\documentclass{article}

\usepackage{graphicx}%
\usepackage{amsmath,amssymb,amsfonts}%
\usepackage{bm,cancel}   
\usepackage{mathtools}
\usepackage{xcolor}%
\usepackage{array}

\usepackage[numbers]{natbib}
\usepackage[a4paper]{geometry}

\newcommand{\lfig}[1]{\label{fig:#1}}
\newcommand{\rfig}[1]{Fig.~\ref{fig:#1}}

\newcommand{\ltab}[1]{\label{tab:#1}}
\newcommand{\rtab}[1]{Table~\ref{tab:#1}}
\newcommand{\lsec}[1]{\label{sec:#1}}
\newcommand{\rsec}[1]{Section~\ref{sec:#1}}
\newcommand{\leqn}[1]{\label{eq:#1}}
\newcommand{\reqn}[1]{(\ref{eq:#1})}

\newcommand{\hh}{\bm{h}}

\newcommand{\uu}{\bm{u}}
\newcommand{\vv}{\bm{v}}
\newcommand{\xx}{\bm{x}}
\newcommand{\dxx}{\dot{\xx}}
\newcommand{\qq}{\bm{q}}
\newcommand{\dqq}{\dot{\qq}}

\newcommand{\thh}{\bm{\theta}}
\newcommand{\dthh}{\dot{\thh}}

\newcommand{\omm}{\bm{\omega}}
\newcommand{\domm}{\dot{\omm}}
\newcommand{\pii}{\bm{\pi}}

\newcommand{\FF}{\bm{F}}

\newcommand{\QQ}{\bm{Q}}
\newcommand{\PP}{\bm{P}}
\newcommand{\Qnc}{Q^{\mathrm{nc}}}
\newcommand{\QQnc}{\QQ^{\mathrm{nc}}}

\newcommand{\XX}{\bm{X}}

\newcommand{\Ia}{\mathord{I'_{\kern-.4ex A}}}
\newcommand{\dIa}{\mbox{$\dot{I}'_{\kern-.4ex A}$}}

\newcommand{\eeZ}{\bm{e}_3}

\newcommand{\tauu}{\bm{\tau}}
\newcommand{\point}[1]{\ensuremath{\mathrm{#1}}}
\newcommand{\varrow}[1]{\,\overrightarrow{\point{#1}}\,}

\newcommand{\clI}{\mathcal{I}}
\newcommand{\clIom}{\mathcal{I}_{\omega}}

\newcommand{\clL}{\mathcal{L}}
\newcommand{\Kine}{\mathcal{T}}
\newcommand{\Pote}{\mathcal{V}}
\newcommand{\Pbk}[1]{\left(#1\right)}
\newcommand{\Bbk}[1]{\left\{#1\right\}}
\newcommand{\Sbk}[1]{\left[#1\right]}
\newcommand{\basis}[1]{\big\langle#1\big\rangle}
\newcommand{\diag}{\mathop{\mathrm{diag}}}

\newcommand{\levi}[3]{\epsilon#1{}#2{}#3}

\makeatletter
\def\@od[#1]#2#3{\frac{d^{#1} #2}{d {#3}^{#1}}}
\def\od{\@ifnextchar[{\@od}{\@od[]}}
\def\pd#1#2{\frac{\partial#1}{\partial#2}}
\def\@lpd[#1]#2#3{\partial^{#1} #2/\partial {#3}^{#1}}
\def\lpd{\@ifnextchar[{\@lpd}{\@lpd[]}}

\makeatother

\newcommand{\OD}[2]{\frac{d'#1}{d#2}}

\newcommand{\bmat}[1]{\begin{bmatrix}#1\end{bmatrix}}

\newcommand{\Skew}[1]{\begin{bmatrix}
        0 & -#1^3 &  #1^2\\
     #1^3 &     0 & -#1^1\\
    -#1^2 &  #1^1 &     0\\
\end{bmatrix}}

\begin{document}

\title{Euler--Lagrange Equations Formulated in Terms of Angular Velocity}
\author{Katsutoshi Yoshida\\
Department of Mechanical Systems Engineering, Utsunomiya University,\\
7-1-2 Yoto, Utsunomiya-shi, Tochigi 321-8585, Japan.\\ 
{\tt yoshidak@a.utsunomiya-u.ac.jp}}

\date{November 7, 2026}


\maketitle

\begin{abstract}%
  We propose a novel variant of the Euler--Lagrange (EL) equation for rigid body rotation, explicitly formulated in terms of orientation parameters $\thh$ and angular velocity $\omm$.
  Variational analysis of the stationarity condition of the action integral for a Lagrangian in terms of $\thh$ and $\omm$ yields a modified EL formulation.
  For pure rotational motion, the proposed equation directly recovers equations of motion fully consistent with the classical Euler equations, bypassing explicit geometry-based torque analysis.
  Furthermore, the framework seamlessly accommodates coupled translational and rotational dynamics, as validated through a spring-suspended symmetrical top, exhibiting exact agreement with the Newton--Euler equations.
  We also identify an inherent discrepancy in the standard EL approach, demonstrating that applying a direct coordinate transformation $\dot\thh\mapsto\omm$ causes the resulting formulation to deviate from both the classical Euler equation and our proposed formulation.
  The proposed framework is expected to provide an intuitive, versatile modeling tool for spatial rigid body dynamics.
\end{abstract}

\section{Introduction}

The Euler--Lagrange (EL) equation is a classical yet modern tool for deriving governing dynamical equations, which is widespread in various fields including physics \cite{Goldstein2011,Landau1976mechanics} and engineering \cite{Williams1996applied,Lamas2025}, and has recently expanded into the field of artificial intelligence in terms of neural networks \cite{Daiz2025ELnet,Motaharifar2026ELnet}.

However, from the perspective of practical applications to spatial rigid body dynamics, 
utilizing the EL equation
\begin{equation}
  \od{}{t}\Pbk{\pd{\clL}{\dot\qq}} 
  - \pd{\clL}{\qq}
  = \bm 0,
\end{equation}
(hereafter referred to as the standard EL equation) 
sometimes poses certain inconveniences.
For example, when dealing exclusively with the rotational motion of a rigid body, 
rather than employing the standard EL equation, 
many practitioners may prefer the classical Euler equations for rotation, 
$I\domm + \omm\times(I\omm)=\tauu$ ($I$: constant inertia tensor, $\omm$: body-fixed angular velocity, $\tauu$: external torque), 
coupled with the kinematic relations $\dthh = B(\thh)\omm$ ($\thh$: orientation).
This is because the standard EL equation requires formulating the Lagrangian $\clL$ in terms of $\qq=\thh$ and $\dqq=\dthh$. 
Consequently, the Lagrangian $\clL$ originally defined in terms of $\omm$ must be rewritten using $(\thh,\dthh)$:
\begin{equation}
  \clL 
  =
  \left. 
  \frac{1}{2}\omm^TI\omm
  \right|_{\omm = B^{-1}(\thh)\dthh}
  = 
  \frac{1}{2}\dthh^T(B(\thh)^{-1})^T I B(\thh)^{-1}\dthh
  ,
\end{equation}
which leads to a computationally complex derivation and a less intuitive structure of the equations of motion compared to the classical Euler equations.
Furthermore, when translational motion $\xx$ coexists with rotational motion for a rigid body, as exemplified by
\begin{equation}
  \clL = \frac{1}{2}\omm^TI\omm + \frac{1}{2}m\dxx^T\dxx + h(\xx,\dxx,\thh,\omm) 
  ,\quad
  \omm = B^{-1}(\thh)\dthh
  ,
\end{equation}
two potential approaches typically emerge.
The first approach is to substitute $\omm = B^{-1}(\thh)\dthh$ into $\clL$ 
and derive the standard EL equations for $(\xx, \dot\xx, \thh, \dot\thh)$, 
thereby accepting the substantial complexity of the resulting formulation.
The second approach is to employ the Newton--Euler equations \cite{Goldstein2011,Williams1996applied}, i.e., 
$m\ddot\xx = \FF$ ($m$: mass, $\FF$: external force) and $I\domm + \omm\times(I\omm)=\tauu$, in combination.
However, this requires an explicit geometry-based force--torque analysis to formulate the coupling terms in $\FF$ and $\tauu$ induced by $h(\xx,\dxx,\thh,\omm)$.
Consequently, neither strategy yields a seamlessly integrated or highly intuitive modeling process.

Therefore, this study explores another approach by representing the Lagrangian $\clL$ explicitly in terms of $\thh$ and $\omm$:
\begin{equation}
  \clL := \clL(\bm\theta, \bm\omega)
  .
\end{equation}
To achieve this, the standard EL equation must be modified accordingly. In this study, we clarify the exact mathematical form of this modification using the calculus of variations and tensor analysis.

Regarding similar concepts, Lamas et al.~\cite{Lamas2025} developed a method to automatically derive the equations of motion for multi-link robots based on a composed vector framework that aggregates translational and angular velocities. 
However, while their work remains within the scope of the standard EL equations, our study takes a distinct approach by explicitly formulating a novel variant of the EL equation via variational principles.

The rest of this paper is organized as follows.
Section 2 provides a brief review of the standard EL equation.
Section 3 derives the proposed EL equation formulated in terms of the rigid body's orientation and angular velocity by analyzing the stationarity condition of the action integral.
Section 4 provides a rigorous comparison of the proposed EL equation with the classical Euler and standard EL equations, and remarks on an inherent discrepancy in the standard EL approach with both the classical Euler equation and our proposed formulation.
Section 5 demonstrates the application of the proposed EL framework to mechanical examples.
Section 6 concludes this paper.

\section{Brief review of the standard EL equation}
\lsec{ELEstd}

First, we briefly review the variational derivation of the standard EL equation \cite{Goldstein2011}.

\subsection{Vector and tensor component notations}

As a preliminary step toward deriving the proposed method, we employ the Einstein summation convention \citep{Lovelock2012Tensor},
where the summation symbol $\sum$ is omitted for products of vector or tensor components with a matching pair of upper and lower indices.

In this study, the components of a vector $\bm v$ are denoted by an upper index $v^{j}$, and the differentiation with respect to a component, $\partial/\partial v^{j}$, is treated as a lower index.
For example, the $i$-th component of the time derivative of a vector-valued function $\bm w = \bm f(\bm v)$ is expressed as follows:
\begin{equation}
  \dot w^i := \Pbk{\od{\bm w}{t}}^i = \pd{f^i}{v^j}\dot v^j
    =\pd{f^i}{v^a}\dot v^a
  .
\end{equation}
The summation index $j$ is referred to as a dummy index, since it can be replaced by any other index (such as $a$ above) without changing the result of the sum. For convenience, dummy indices are often changed during operations such as substitution or matching terms.

In addition, this study adopts the following notations, which are also used in the above equation. The time derivative is represented by a dot, i.e., $\dot x := dx/dt$. The components of a vector $\bm v$ are written as $v^j$ or $(\bm v)^j$, and the components of a matrix $A$ are written as $a^i_j$ ($a^{\text{row}}_{\text{column}}$) or $(A)^i_j$. Conversely, a vector with components $v^j$ is denoted by $\bm v = [v^j]$, and a matrix with components $a^{i}_{j}$ by $A = [a^{i}_{j}]$.

\subsection{Variational method}

In Lagrangian mechanics \cite{Goldstein2011,Landau1976mechanics}, it is stated that the realized trajectory $\qq(t)$ satisfies the stationarity condition $\delta\clI=0$ of the action integral $\clI$.

Specifically, for a mechanical system with generalized coordinates $\qq=[q^i]$ ($1\le i\le N$) and 
a time-dependent Lagrangian $\clL(\qq,\dqq,t)$, the action integral $\clI$ is defined as
\begin{equation}
  \clI(\qq,\dqq) := \int_{t_0}^{t_1}\clL(\qq,\dqq,t)dt.
\end{equation}
Considering an arbitrary variation $\delta\qq=[\delta q^i]$ under the boundary conditions:
\begin{equation}
  \delta\qq(t_0) = \delta\qq(t_1) = \bm 0,
  \leqn{orig:delta}
\end{equation}
the resulting variation of the action integral becomes
\begin{equation}
  \delta\clI(\qq,\dqq) := \clI(\qq+\delta\qq,\dqq+\delta\dqq) - \clI(\qq,\dqq)
  =: \int_{t_0}^{t_1} \delta \clL(\qq,\dqq,t) dt.
  \leqn{orig:deltaI}
\end{equation}
The stationarity condition is then given by $\delta\clI(\qq,\dqq)=0$ \cite{Goldstein2011}.

\subsection{Derivation of the standard EL equation}

The trajectory $\qq(t)$ that satisfies the stationarity condition $\delta\clI(\qq,\dqq)=0$ is obtained as follows \cite{Goldstein2011}. First, expanding the variation of the Lagrangian $\delta\clL$ in \reqn{orig:deltaI} and neglecting second- and higher-order terms of $\delta\qq$ yields the following equation:
\begin{equation}
  \delta\clL(\qq,\dqq,t)
  :=
  \clL(\qq+\delta\qq,\dqq+\delta\dqq,t) - \clL(\qq,\dqq,t)
  \approx \pd{\clL}{q^j}\delta q^j + \pd{\clL}{\dot q^j}\delta \dot q^j 
  \leqn{orig:dI:0}
\end{equation}
Substituting \reqn{orig:dI:0} back into $\delta \clI$ in \reqn{orig:deltaI}, applying integration by parts, and using the boundary conditions for the variation in \reqn{orig:delta}, the stationarity condition becomes
\begin{align}
  0 = 
  \delta\clI(\qq,\dqq)
  & 
  = \int_{t_0}^{t_1} \delta \clL(\qq,\dqq,t) dt
  \notag\\&
  = \int_{t_0}^{t_1} 
      \pd{\clL}{q^j}\delta q^j
    dt 
    + \int_{t_0}^{t_1}
      \pd{\clL}{\dot q^j}\delta \dot q^j 
    dt
  \notag\\&
  = \int_{t_0}^{t_1} 
      \pd{\clL}{q^j}\delta q^j
    dt
    - \int_{t_0}^{t_1}
    \od{}{t}\Pbk{\pd{\clL}{\dot q^j}}\delta q^j 
    dt
    +
    \Sbk{\pd{\clL}{\dot q^j}\delta q^j}_{t_0}^{t_1} 
  \notag\\&
  = \int_{t_0}^{t_1}
    \Bbk{
      \pd{\clL}{q^j}
      -\od{}{t}\Pbk{\pd{\clL}{\dot q^j}}
    }
      \delta q^j
    dt.
  \leqn{orig:dI:1}
\end{align}
For \reqn{orig:dI:1} to hold for arbitrary variation $\delta q^j$, the following equation is required \cite{Goldstein2011}:
\begin{equation}
  \od{}{t}\Pbk{\pd{\clL}{\dot q^j}} 
  - \pd{\clL}{q^j}
  = 0
  .
\end{equation}
This provides the standard force-free EL equation. 

Furthremore, considering the non-conservative generalized force $\Qnc_j$ associated with the generalized coordinates $q^j$, D'Alembert's principle holds \cite{Goldstein2011}:
\begin{equation}
      \Bbk{
        \od{}{t}\Pbk{\pd{\clL}{\dot q^j}} 
        - \pd{\clL}{q^j}
        - \Qnc_j(\qq)
    }
    \delta q^j
    =0
    .
    \leqn{D'Alenbert}
\end{equation}
Therefore, since $\delta q^j$ is arbitrary, the following equation is required:
\begin{equation}
  \od{}{t}\Pbk{\pd{\clL}{\dot q^j}} 
  - \pd{\clL}{q^j}
  = \Qnc_j(\qq)
  ,
  \leqn{ELE:original}
\end{equation}
or equivalently in column-vector form,
\begin{equation}
  \od{}{t}\Pbk{\pd{\clL}{\dot\qq}} 
  - \pd{\clL}{\qq}
  = \QQnc(\qq)
  ,
  \leqn{ELE:original:vector}
\end{equation}
where $\partial\clL/\partial\qq$ denotes the column-vector-valued gradient with respect to the vector $\qq$. These equations recover the standard forced EL equations.

\section{Angular velocity representation of the EL equation}
\lsec{ELEom}

Focusing on the rotational motion of a spatial rigid body, we derive the stationarity condition of the action integral for the Lagrangian formulated as a function of orientation $\thh$ and angular velocity $\omm$, using a variational method similar to that in \rsec{ELEstd}.

\subsection{Kinematic assumptions}

We choose the inertial frame $XYZ$ as the reference frame. Furthermore, we introduce a body-fixed frame $xyz$ aligned with the principal axes of inertia of the rigid body, and define the reference orientation of the rigid body as the orientation where the frame $xyz$ coincides with $XYZ$.
Hereafter, primed and unprimed symbols denote quantities in the reference frame and the body-fixed frame, respectively. 

\subsubsection{Orientation representation}

Let $\basis{\uu'_i}:=\basis{\uu'_1,\uu'_2,\uu'_3}$ denote the basis of the reference frame $XYZ$, where $\uu'_i$ are the basis vectors.
Similarly, let $\basis{\uu_i}$ denote the basis of the body-fixed frame $xyz$, and represent the orientation of $\basis{\uu_i}$ relative to $\basis{\uu'_i}$ by a rotation matrix $R$. That is,
\begin{equation}
    \basis{\uu_i} = \basis{R\uu'_i}
    .
    \leqn{trans:basis}
\end{equation}
By parameterizing the matrix as $R=[R^i_j]=R^i_j(\thh)$, we adopt the parameters $\thh:=[\theta^i]$ ($1\le i \le N$) as the orientation representation of the rigid body. 
In this study, we consider either Euler angles ($N=3$) or Euler parameters ($N=4$) \citep{Shabana2005MBD} for $\thh$.
Note that since the rotation matrix is orthogonal, the following relation holds:
\begin{equation}
  R^{-1} = R^T
  .
\end{equation}
By using this, any vector $\vv'$ in the reference frame is transformed into the corresponding $\vv$ in the body-fixed frame as follows:
\begin{equation}
  \vv' = R\vv\quad\text{or}\quad \vv = R^T\vv'
  .
  \leqn{trans:vector}  
\end{equation}

\subsubsection{kinematic relations}
\lsec{kinematic relation}

In the body-fixed frame, the angular velocity vector $\omm = [\omega^i]$ relates to the rotation matrix $R =[R^i_j(\thh)]$ as follows \cite{Shabana2005MBD}:
\begin{equation}
  \dxx = \omm\times\xx = [\omm\times]\xx
  ,\quad 
  [\omm\times] := R^T\dot{R} = \Skew{\omega}
  ,
  \leqn{cross:matrix:body}
\end{equation}
where $\xx$ and $[\omm\times]$ denote a vector and a skew-symmetric cross-product matrix, respectively.
The body-fixed representation of angular velocity ensures a constant inertia tensor and is widely utilized across many fields; this convention is adopted throughout this study.

Here, introducing the Levi-Civita symbol \citep{Riley2006math}
\begin{equation}
  \epsilon_{ijk}
  =
    \begin{cases}
      1,   & \text{if $ijk = 123, 231, 312$},\\
      -1,  & \text{if $ijk = 132, 321, 213$},\\
      0,   & \text{otherwise},
    \end{cases}
  \leqn{Levi}
\end{equation}
(where, for simplicity, no distinction is made between upper and lower indices), the cross product can be expressed as 
$(\omm\times\xx)^i=\levi{^i}{_{jk}}{}\omega^jx^k$. Therefore, the formula \reqn{cross:matrix:body} can be written in the following compact notation:
\begin{equation}
  \dot{x}^i 
  = \levi{^i}{_{jk}}{}\omega^j x^k 
  = {R^T}^i_j \dot{R}^j_k x^k 
  = {R^T}^i_j \pd{R^j_k}{\theta^l}\dot{\theta}^l  x^k
  ,\quad
  [\omm\times]^i_k
  =
  {R^T}^i_b\pd{R^b_k}{\theta^l}\dot{\theta}^l  
  = 
  \levi{^i}{_{bk}}\omega^b 
  .
  \leqn{om:formula}
\end{equation}
To solve for $\omega^j$, we multiply both sides by $\epsilon_{ijk}/2$ and apply the identity for the Levi-Civita symbol \citep{Riley2006math}:
\begin{equation}
  2\delta^a_b
  =\levi{^a}{_{jk}}{}\levi{_b}{^{jk}}{}
  =\levi{_i}{^a}{_k}\levi{^i}{_b}{^k}
  =\levi{_{ij}}{^a}\levi{^{ij}}{_b}
  = \cdots
  ,
  \leqn{Levi_delta}
\end{equation}
which yields the following equation:
\begin{align}
  \frac{1}{2}
  \levi{_i}{^{jk}}{}
  \levi{^i}{_{bk}}{}\omega^b
  &
  = \delta^j_b\omega^b
  \notag\\
  &
  =
  \omega^j
  =
  \frac{1}{2}
  \levi{_i}{^{jk}}{}
  {R^T}^i_b\pd{R^b_k}{\theta^l}\dot{\theta}^l  
  =
  \frac{1}{2}
  \levi{_a}{^{jc}}{}
  {R^T}^a_b\pd{R^b_c}{\theta^i}\dot{\theta}^i
  .
  \leqn{A:tensor}
\end{align}
That is, the components of the angular velocity vector, $\omega^j$, can be explicitly written in terms of the rotation matrix and its partial derivatives:
\begin{equation}
  \begin{dcases}
    \omega^j
    =
    A^j_i
    \dot\theta^i
    ,\;
    A^j_i 
    = 
    \frac{1}{2}
    \levi{_a}{^{jc}}{}
    {R^T}^a_b\pd{R^b_c}{\theta^i}
    ,\;
    1\le i \le N,\;1\le j \le M
    ,
    \\
    \omm = A\dthh
    ,\;
    A = [A^j_i] = \left[
      \frac{1}{2}
      \levi{_a}{^{jc}}{}
      {R^T}^a_b\pd{R^b_c}{\theta^i}
    \right]
    ,
    \qquad \text{(Column-vector form)}
  \end{dcases}
  \leqn{A}
\end{equation}
which defines the kinematic relations \citep{Shabana2005MBD} mapping $\dthh$ to $\omm$; the upper and lower rows represent the tensor and vector forms, respectively.

When the orientation representation $\thh$ consists of Euler parameters \citep{Shabana2005MBD}, the inverse of $A$ does not exist because $\dim \omm = 3 < \dim \thh = 4$. Therefore, the inverse relationship is expressed using the right pseudoinverse $B$ of $A$:
\begin{equation}
  \begin{dcases}
    \dot\theta^i
    =
    B^i_j
    \omega^j
    ,\;
    A^j_iB^i_k 
    = 
    \delta^{j}_{k}
    ,\;
    1\le i \le N,\;1\le j,k \le M
    ,
    \\
    \dthh = B\omm
    ,\;
    AB = \text{Identity matrix} 
    ,
    \qquad \text{(Column-vector form)}
  \end{dcases}
  \leqn{B}
\end{equation}
which provides the inverse kinematic relations mapping $\omm$ to $\dthh$.
The symbols used above are summarized in \rtab{Symbols}.

\begin{table}[t]
  \centering
  \caption{Symbols for rigid body rotation in the inertial reference frame $XYZ$.}
  \ltab{Symbols}
  \begin{tabular}{lll}
    \hline
    Symbol & Definition & Remarks 
    \\\hline
    $\theta^i$ & Orientation & $1\le i\le N=3$ for Euler angles, \\
    && $1\le i\le N=4$ for Euler parameters
    \\\hline
    $\omega^j$ & Angular velocity components &
    $1\le j\le M = 3$
    \\ & in the body-fixed frame &
    \\\hline
    $\omega^j = A^j_i\dot\theta^i$ & Kinematic relation &
    $B^i_j:=$ Pseudoinverse of $A^j_i$  
    \\
    $\dot\theta^i := B^i_j\omega^j$ & Inverse kinematic relation &
    such that $A^j_i B^i_k= \delta^j_k$
    \\\hline
  \end{tabular}
\end{table}

\subsubsection{Variation of angular velocity}

Due to the analogy between the derivative and variational operators, 
the relations \reqn{cross:matrix:body} and \reqn{A} for the body-fixed frame 
yield the following variational relations \cite{Shabana2005MBD}:
\begin{equation}
  [\delta\pii\times] = R^T\delta{R}
  ,\quad
  \delta\pii = A\delta\thh
  ,
\end{equation}
where $\delta\pii$ is the infinitesimal rotation vector and $\delta\thh$ is the variation of the orientation.
By utilizing $[\delta\pii\times]$, the commutativity of variation and differentiation, and the commutator product,
the variation of the cross-product matrix \reqn{cross:matrix:body} is obtained as
\begin{align}
  \delta[\omm\times] 
  &= \delta(R^T\dot{R})
  \notag\\&
  = \delta{R}^T\dot{R} + R^T(\delta{R})\dot{} 
  \notag\\&
  = \Pbk{R[\delta\pii\times]}^T{R[\omm\times]} + R^T(R[\delta\pii\times])\dot{}
  \notag\\&
  = -[\delta\pii\times][\omm\times] + R^T\dot{R}[\delta\pii\times] + R^TR ([\delta\pii\times])\dot{}
  \notag\\&
  = ([\delta\pii\times])\dot{} + [\omm\times][\delta\pii\times] - [\delta\pii\times][\omm\times]
  \notag\\&
  = [(\delta\pii)\dot{}\times] + [(\omm\times\delta\pii)\times]
  ,
\end{align}
where $(\cdot)\dot{} := d(\cdot)/dt$ denotes the time derivative. This yields the variation of the angular velocity in component form and in column-vector form, respectively, as
\begin{equation}
  \begin{cases}
    \delta{\omega}^j = (\delta{\pi}^j)\dot{} + \levi{^j}{_{ab}}{}\omega^a\delta{\pi}^b
    ,
    \\
    \delta{\omm} = (\delta{\pii})\dot{} + \omm\times\delta{\pii}
    .
    \qquad \text{(Column-vector form)}
  \end{cases}
\end{equation}
Substituting $\delta{\pi}^j = A^j_i\delta{\theta}^i$ into this yields
\begin{align}
  \delta{\omega}^j &
  = \dot{A}^j_i\delta{\theta}^i + A^j_i\delta{\dot\theta}^i + \levi{^j}{_{ab}}{}\omega^aA^b_i\delta{\theta}^i
  \notag\\&
  = \Pbk{\dot{A}^j_i + \levi{^j}{_{ab}}{}\omega^aA^b_i}\delta{\theta}^i + A^j_i\delta{\dot\theta}^i
  .
  \leqn{delta:om}
\end{align}

\subsection{Variational method}

Consider a Lagrangian $\clL$ defined in terms of the orientation $\thh$, the body-fixed angular velocity vector $\omm$, and time $t$. Its corresponding action integral $\mathcal{I}_\omega$ is defined as follows:
\begin{align}
  &
  \clL = \clL(\thh, \omm, t) = \Kine(\thh, \omm, t) - \Pote(\thh, t)
  ,
  \\
  &
  \clIom = \clIom(\thh, \omm) := \int_{t_0}^{t_1}\clL(\thh, \omm, t)dt
  ,
  \leqn{om:I}
\end{align}
where $\Kine$ and $\Pote$ denote the kinetic and potential energies of the rigid body, respectively.

For arbitrary variations $\delta\thh=[\delta\theta^i]$ subject to the boundary conditions
\begin{equation}
  \delta\thh(t_0) = \delta\thh(t_1) = \bm{0},
  \leqn{delta:theta}
\end{equation}
the variation of the action integral is given by
\begin{align}
  \delta\clIom(\thh,\omm) 
  &:= \clIom(\thh+\delta\thh,\omm+\delta\omm) - \clIom(\thh,\omm)
  \notag\\&
  = \int_{t_0}^{t_1} \delta \clL(\thh,\omm,t) dt.
  \leqn{om:deltaI}
\end{align}
Then, the stationarity condition for the proposed formulation is given by $\delta\clIom(\thh,\omm)=0$.

\subsection{Derivation of the proposed EL equation}
\lsec{ProposedELE}

As in \rsec{ELEstd}, we expand the variation of the Lagrangian in \reqn{om:deltaI}. 
Neglecting higher-order terms of $\delta\theta^i$ and $\delta\omega^j$, the expansion yields
\begin{align}
  \delta \clL(\thh,\omm,t)
  &=
  \clL(\thh+\delta\thh,\omm+\delta\omm,t) - \clL(\thh,\omm,t)
  \notag \\&
  =
  \pd{\clL}{\theta^i} \delta\theta^i
  + 
  \pd{\clL}{\omega^j} \delta\omega^j
  \notag \\&
  =
  \pd{\clL}{\theta^i} \delta\theta^i
  + 
  \pd{\clL}{\omega^j} 
  \Bbk{\Pbk{\dot{A}^j_i + \levi{^j}{_{ab}}{}\omega^aA^b_i}\delta{\theta}^i + A^j_i\delta{\dot\theta}^i}
  \notag \\&
  =
  \Bbk{
    \pd{\clL}{\theta^i}
    +
    \pd{\clL}{\omega^j}
    \Pbk{\dot{A}^j_i + \levi{^j}{_{ab}}{}\omega^aA^b_i}
  }
  \delta\theta^i
  + 
  \pd{\clL}{\omega^j}
  A^j_i
  \delta{\dot\theta}^i 
  .
  \leqn{om:dL}
\end{align}
Integrating \reqn{om:dL} by parts and applying the variational boundary conditions \reqn{delta:theta} yields the stationarity condition
\begin{align}
  0 
  &=
  \delta\clIom(\thh,\omm)
  \notag \\
  &=
  \int_{t_0}^{t_1} \delta \clL(\thh,\omm,t) dt
  \notag \\
  &=
  \int_{t_0}^{t_1} 
  \Bbk{
    \pd{\clL}{\theta^i}
    +
    \pd{\clL}{\omega^j}
    \Pbk{\dot{A}^j_i + \levi{^j}{_{ab}}{}\omega^aA^b_i}
  }
  \delta\theta^i
  dt
  -
  \int_{t_0}^{t_1} 
  \od{}{t}\Pbk{\pd{\clL}{\omega^j}A^j_i} \delta\theta^i
  dt
  + \Sbk{
    \pd{\clL}{\omega^j}
    A^j_i
    \delta\theta^i
  }_{t_0}^{t_1}
  \notag \\
  &=
  \int_{t_0}^{t_1} 
  \Bbk{
    \pd{\clL}{\theta^i}
    +
    \pd{\clL}{\omega^j}
    \Pbk{\cancel{\dot{A}^j_i} + \levi{^j}{_{ab}}{}\omega^aA^b_i}
    -
    \od{}{t}\Pbk{\pd{\clL}{\omega^j}}A^j_i
    -
    \cancel{\pd{\clL}{\omega^j}\dot{A}^j_i}
  }
  \delta\theta^i
  dt
  .
\end{align}
For this to hold for an arbitrary variation $\delta\theta^i$, the following equation must be satisfied:
\begin{equation}
  \od{}{t}\Pbk{\pd{\clL}{\omega^j}}A^j_i
  -
  \levi{^j}{_{ab}}{}\omega^aA^b_i\pd{\clL}{\omega^j} 
  -
  \pd{\clL}{\theta^i}
  =0
  .
\end{equation}
Furthermore, considering the non-conservative generalized force $\Qnc_j(\thh)$ associated with the generalized coordinates $\theta^j$, D'Alembert's principle \cite{Goldstein2011} can be expressed as
\begin{equation}
  \Bbk{
    \od{}{t}\Pbk{\pd{\clL}{\omega^j}}A^j_i
    -
    \levi{^j}{_{ab}}{}\omega^aA^b_i\pd{\clL}{\omega^j} 
    -
    \pd{\clL}{\theta^i}
    -\Qnc_i(\thh)
  }
  \delta{\theta}^i
  =0,
\end{equation}
which yields
\begin{equation}
  \od{}{t}\Pbk{\pd{\clL}{\omega^j}}A^j_i
  -
  \levi{^j}{_{ab}}{}\omega^aA^b_i\pd{\clL}{\omega^j} 
  -
  \pd{\clL}{\theta^i}
  =\Qnc_i(\thh)
  .
  \leqn{modELE:tensor:A}
\end{equation}

The proposed angular representation of the EL equation is derived
by post-multiplying both sides of \reqn{modELE:tensor:A} by the right pseudoinverse $B^i_k$:
\begin{equation}
  \od{}{t}\Pbk{\pd{\clL}{\omega^k}}
  +
  \levi{_{ka}}{^j}{}\omega^a\pd{\clL}{\omega^j} 
  -
  \pd{\clL}{\theta^i}B^i_k
  =\Qnc_i(\thh)B^i_k
  \quad(\because\levi{^j}{_{ak}}{} = -\levi{_{ka}}{^j}{})
  ,
\end{equation}
which is further represented in column-vector form,
\begin{equation}
  \od{}{t}\Pbk{\pd{\clL}{\omm}}
  +
  \omm\times \pd{\clL}{\omm}
  -
  B^T\pd{\clL}{\thh}
  =
  B^T\QQnc(\thh)
  ,
  \leqn{modELE:vector:body}
\end{equation}
where $\partial\clL/\partial\xx$ denotes the column-vector-valued gradient with respect to the vector $\xx$.
The total solution $\XX(t) = (\thh(t)^T,\omm(t)^T)^T$ for the rotational motion can be obtained by solving \reqn{modELE:vector:body} simultaneously with the inverse kinematic relation derived in \reqn{B}.

\subsection{General formulation}

The above formulation can be straightforwardly generalized to Lagrangians 
defined in terms of the orientation--velocity pair $(\thh, \omm)$ 
alongside other generalized coordinates and velocities $(\qq,\dot\qq)$.
The variation of such a Lagrangian yields the linear combination of the variations of the variables:
\begin{align}
  \delta \clL(\thh,\omm,\qq,\dot\qq,t)
  &=
  \clL(\thh+\delta\thh,\omm+\delta\omm,\qq+\delta\qq,\dot\qq+\delta\dot\qq,t) - \clL(\thh,\omm,\qq,\dot\qq,t)
  \notag \\
  &=
  \pd{\clL}{\theta^i} \delta\theta^i
  + 
  \pd{\clL}{\omega^j} \delta\omega^j
  +
  \pd{\clL}{q^k} \delta q^k
  + 
  \pd{\clL}{\dot{q}^l} \delta \dot{q}^l
  ,
\end{align}
where the corresponding variation of the action integral can be decomposed into two distinct parts:
\begin{equation}
  \delta\clI 
  = 
  \delta\clI_{\omega} + \delta\clI_q
  := \int_{t_0}^{t_1}\Pbk{
    \pd{\clL}{\theta^i} \delta\theta^i
    + 
    \pd{\clL}{\omega^j} \delta\omega^j
  }dt
  +
  \int_{t_0}^{t_1}\Pbk{
    \pd{\clL}{q^k} \delta q^k
    + 
    \pd{\clL}{\dot{q}^l} \delta \dot{q}^l
  }dt
  .
\end{equation}
This decomposition makes it possible to decouple the original stationarity condition $\delta\clI=0$ into
the independent subconditions $\delta\clI_{\omega} = 0$ and $\delta\clI_q = 0$.
 
Then, applying 
the proposed derivation in \rsec{ELEom} to $\delta\clI_{\omega} = 0$ and
the standard derivation in \rsec{ELEstd} to $\delta\clI_q = 0$ 
yields an independent set of equations comprising 
the proposed EL equation \reqn{modELE:vector:body} for $(\thh, \omm)$
and 
the standard EL equation \reqn{ELE:original:vector} for $(\qq,\dot\qq)$:
\begin{equation}
  \begin{dcases}
    \od{}{t}\Pbk{\pd{\clL}{\omm}}
    +
    \omm\times \pd{\clL}{\omm}
    -
    B^T\pd{\clL}{\thh}
    =
    B^T\QQnc(\thh)
    , 
    \\
    \od{}{t}\Pbk{\pd{\clL}{\dot\qq}}
    -
    \pd{\clL}{\qq}
    =
    \QQnc(\qq)
    .
  \end{dcases}
  \leqn{ELE:general}
\end{equation}
This set of equations provides a novel counterpart to the classical Newton--Euler equations \cite{Goldstein2011,Williams1996applied}, as demonstrated in \rsec{symtop_spring}.

\section{Comparison with conventional equations}

To verify the proposed EL equation, this section considers the rotational motion of a rigid body 
subject solely to a potential energy $V(\thh)$ depending on its orientation $\thh$, 
in comparison with the classical Euler and standard EL equations. 
Let $I$ and $\omm$ denote the body-fixed inertia tensor and angular velocity vector, respectively.

\subsection{Consistency with the classical Euler equation}
\lsec{classical:Euler}

The classical Euler equation \cite{Goldstein2011} is given by
\begin{equation}
  I\domm + \omm\times(I\omm) = \tauu
  ,
\end{equation}
where $\tauu$ denotes the external torque vector about the body-fixed axes $xyz$.
For this system, the virtual work $\delta W$ done by the potential force can be expressed in terms of either 
the infinitesimal rotation vector $\delta\pii$ 
or 
the orientation variation $\delta\thh$, as \cite{Goldstein2011}
\begin{equation}
  \delta W 
  = \delta\pii^T \tauu
  = \Pbk{\delta\thh^TA^T} \tauu
  ,\quad
  \delta W
  = - \delta\thh^T\Bbk{\pd{\Pote(\thh)}{\thh}}
  ,
  \leqn{comparison:Euler:force:begin}
\end{equation}
which yields the relation
\begin{equation}
  A^T\tauu
  = 
  - \pd{\Pote(\thh)}{\thh} 
  .
\end{equation}
Pre-multiplying both sides by $B^T$ yields the torque $\tauu$ generated from the potential $\Pote(\thh)$, namely, 
\begin{equation}
  \tauu = B^TA^T\tauu= - B^T\pd{\Pote(\thh)}{\thh}
  .
  \leqn{comparison:Euler:force:end}
\end{equation}
Therefore, the resulting classical Euler equation for this problem is given by
\begin{equation}
  I\domm + \omm\times(I\omm) = - B^T\pd{\Pote(\thh)}{\thh}
  .
  \leqn{comparison:Euler}
\end{equation}

On the other hand, this problem can also be described by the proposed EL equation \reqn{modELE:vector:body} associated with the following Lagrangian and generalized force: 
\begin{equation}
  \clL = \frac{1}{2}\omm^TI\omm - \Pote(\thh)
  ,\quad
  \QQnc = \bm0
  .
  \leqn{comparison:L:om}
\end{equation}
Substituting \reqn{comparison:L:om} into \reqn{modELE:vector:body} yields
\begin{align}
  &\od{}{t}\Pbk{\pd{\clL}{\omm}}
  +
  \omm\times \pd{\clL}{\omm}
  -
  B^T\pd{\clL}{\thh}
  \notag\\
  &=
  I\domm + \omm\times(I\omm) + B^T\pd{\Pote(\thh)}{\thh} = \bm 0
  ,
  \leqn{comparison:ELEom}
\end{align}
which is identical to the classical Euler equation \reqn{comparison:Euler}.

The above comparison successfully demonstrates that 
our proposed EL equation \reqn{modELE:vector:body} combines the advantages of both the classical Euler and standard EL equations.
First, it immediately yields the equations of motion equivalent to the classical Euler equation.
Second, force analyses such as those from \reqn{comparison:Euler:force:begin} to \reqn{comparison:Euler:force:end} are no longer required; as with the standard EL equation, simply defining the potential energy automatically handles the necessary coordinate transformations.

\subsection{Remarks on difference from the standard EL equation}

We apply the same comparison in terms of $(\thh,\omm)$ to the standard EL equation
\reqn{ELE:original:vector},
\begin{equation}
  \od{}{t}\Pbk{\pd{\clL}{\dthh}} - \pd{\clL}{\thh} = \QQnc(\thh)
  ,
  \leqn{ELEstd:dthh}
\end{equation}
via a coordinate transformation from $(\thh,\dthh)$ to $(\thh,\omm)$.
In short, unlike the proposed EL equation (verified in \rsec{classical:Euler}), 
consistency with the classical Euler equation does not hold for the standard EL equation under such a coordinate transformation.

To demonstrate this, we apply the transformation not to the Lagrangian but directly to the equation \reqn{ELEstd:dthh} 
using the relation
\begin{equation}
  \pd{\clL}{\dot{\theta}^i}
  =
  \pd{\omega^j}{\dot{\theta}^i}\pd{\clL}{\omega^j}
  =
  A^j_i\pd{\clL}{\omega^j}
  \quad\text{($\because\;\omega^j = A^j_i\dot{\theta}^i$)}
  ,
\end{equation}
or equivalently in column-vector form,
\begin{equation}
  \pd{\clL}{\dthh}
  = A^T\pd{\clL}{\omm}
  .
  \leqn{trans:dth->om}
\end{equation}
Substituting \reqn{trans:dth->om} into \reqn{ELEstd:dthh} and pre-multiplying both sides by $B^T$ yields
\begin{equation}
  \od{}{t}\Pbk{\pd{\clL}{\omm}} + (\dot{A}B)^T \pd{\clL}{\omm} - B^T\pd{\clL}{\thh} = B^T\QQnc(\thh)
  ,
  \leqn{ELEstd:om}
\end{equation}
which provides the $(\thh,\omm)$-representation of the standard EL equation \reqn{ELEstd:dthh}.

\begin{table}
  \newcommand{\mypadding}{\rule[-2.7ex]{0pt}{7.2ex}}
  \centering
  \caption{Comparison of the proposed EL equation with the standard EL equation, its $(\thh,\omm)$-representation, and the classical Euler equation.
  The standard EL equation represented in $(\thh, \omm)$ fails to exhibit consistency with the classical Euler equation due to $(\dot{A}B)^T \neq [\omm\times]$,
  unlike the proposed EL equation.}
  \ltab{comparison}
  \newcommand{\dblrow}[2]{$\vcenter{\vfil\hbox{\sf #1\hss}\kern.5ex\hbox{\sf #2\hss}\vfil}$}
  \begin{tabular}{l@{}cl}
    \hline
    \multicolumn{1}{c}{Approach} & \multicolumn{1}{c}{Coordinates} & \multicolumn{1}{c}{Equation}\\ \hline
    \dblrow{(1) Standard}{EL equation}
    & $\thh,\dthh$ &
    \mypadding
    \begin{math}
      \displaystyle
      \od{}{t}\Pbk{\pd{\clL(\thh,\dthh)}{\dthh}} - \pd{\clL(\thh,\dthh)}{\thh} = \QQnc
    \end{math}
    \\\cline{2-3}
    & $\thh,\omm$ &
    \mypadding
    \begin{math}
      \displaystyle
      \od{}{t}\Pbk{\pd{\clL(\thh,\omm)}{\omm}} + (\dot{A}B)^T \pd{\clL(\thh,\omm)}{\omm} - B^T\pd{\clL(\thh,\omm)}{\thh} = B^T\QQnc
    \end{math}
    \\\hline\hline
    \dblrow{(2) Proposed}{EL equation}
    & $\thh,\omm$ &
    \mypadding
    \begin{math}
      \displaystyle
      \od{}{t}\Pbk{\pd{\clL(\thh,\omm)}{\omm}} \;\;\;+ \omm\times \pd{\clL(\thh,\omm)}{\omm} - B^T\pd{\clL(\thh,\omm)}{\thh} = B^T\QQnc
    \end{math}
    \\\hline
    \dblrow{(3) Classical}{Euler equation}
    & $\thh,\omm$ &
    \mypadding
    \begin{math}
      \displaystyle
      \od{}{t}\Pbk{I\omm} + \omm\times \Pbk{I\omm} - B^T\pd{(-\Pote)}{\thh} = \bm 0
    \end{math}
    \\\hline
  \end{tabular}
\end{table}

\rtab{comparison} compares the proposed EL equation \reqn{modELE:vector:body} with 
the standard EL equation \reqn{ELEstd:dthh}, its $(\thh,\omm)$-representation \reqn{ELEstd:om}, and the classical Euler equation \reqn{comparison:Euler}.
This comparison clearly demonstrates that the standard EL equation \reqn{ELEstd:om} exhibits the discrepancy
\begin{equation}
  \Delta_{\text{EL}} := (\dot{A}B)^T - [\omm\times]
  \leqn{difference} 
\end{equation}
relative to the proposed EL equation \reqn{modELE:vector:body}.
The same discrepancy appears relative to the classical Euler equation \reqn{comparison:Euler}, 
as substituting $\clL$ and $\QQnc$ from \reqn{comparison:L:om} into \reqn{ELEstd:om} yields
\begin{equation}
  I\domm 
  + (\dot{A}B)^TI\omm 
  + B^T\pd{\Pote(\thh)}{\thh} 
  = \bm0
  .
  \leqn{comparison:ELEstd}
\end{equation}

Importantly, the equality $\Delta_{\text{EL}} = O$ does not hold in general, as demonstrated below.
A counterexample is provided by expressing $(\dot{A}B)^T$ in component form as
\begin{equation}
  (\dot{A}B)^T{}^i_k = (\dot{A}B)^{k}_{i}
  =
  \pd{A{}^k_l}{\theta^m}B^l_i\dot{\theta}^m
  =
  \pd{A^k_l}{\theta^m}B^l_iB^m_j\omega^j
  =
  \beta_{ij}{}^k\omega^j
  ,\;
  \beta_{ij}{}^k
  :=
  B^l_iB^m_j\pd{A^k_l}{\theta^m}
  .
  \leqn{dotABT}
\end{equation}
Hence, the equality $\Delta_{\text{EL}}=O$ is equivalent to 
$\beta_{ij}{}^k=\levi{^i}{_{jk}}{}$ in terms of value, 
which yields 
\begin{align}
  (\dot{A}B)^T{}^i_k
  &=(\dot{A}B){}^k_i = \beta_{ij}{}^k\omega^j
  = \levi{^i}{_{jk}}{}\omega^j
  = [\omm\times]^i_k.
\end{align}

Specifically, for the Euler parameters:
\begin{equation}
  \thh = \hh = (h^0, h^1, h^2, h^3)^T = (\cos\vartheta, a^1\sin\vartheta, a^2\sin\vartheta, a^3\sin\vartheta)^T
  ,
\end{equation}
where $\vartheta$ and $\bm{a} = [a^i]$ ($\bm{a}^T\bm{a}=1$) denote the rotation angle and rotation axis, respectively.
The corresponding kinematic matrices are given as follows \cite{Shabana2005MBD}:
\begin{gather}
  A = 2
  \begin{bmatrix}
    -h^1&  h^0&  h^3& -h^2\\
    -h^2& -h^3&  h^0&  h^1\\
    -h^3&  h^2& -h^1&  h^0\\   
  \end{bmatrix}
  ,\quad
  B = \frac{1}{4}A^T
  .
\end{gather}
By substituting these matrices into \reqn{dotABT}, it can be proved that
\begin{equation}
  \beta_{ij}{}^k=\frac{1}{2}\levi{^i}{_{jk}}{}
  \quad\text{or}\quad
  (\dot{A}B)^T = \frac{1}{2}[\omm\times]
  ,
\end{equation}
indicating that the equality is violated, i.e., 
$\Delta_{\text{EL}} = (\dot{A}B)^T - [\omm\times] = -\frac{1}{2}[\omm\times] \neq O$.

Euler angles also yield counterexamples.
For instance, for the ZXZ Euler angles $\thh=(\theta_1,\theta_2,\theta_3)^T$, it can be proved that $\beta_{12}{}^3 = \sin^2\theta_3 \neq 1 = \epsilon^1{}_{23}$, confirming that $\Delta_{\text{EL}} \neq O$.
Similarly, other Euler angle conventions also yield counterexamples, whose detailed derivations are omitted here.

The above verification suggests that, regarding the rotational motion of a rigid body, the standard EL equation fails to match both the proposed EL equation and the classical Euler equation. While this discrepancy is of substantial physical interest, a detailed analysis lies beyond the scope of the present study and will be addressed in future work.

\newpage
\section{Mechanical examples}

The practicality of the proposed EL formulation is demonstrated
by applying it to two types of spinning tops: one with a fixed pivot and the other with a spring support.
The former serves to recover the rigorous discussion of a pure rotational problem in \rsec{classical:Euler}, 
while the latter validates the formulation by reproducing the Newton--Euler equations for the coupled translational and rotational motion.

\subsection{Symmetrical top with a fixed pivot}
\lsec{symtop_fixed}

Following the classic formulation \cite{Goldstein2011,Landau1976mechanics}, we consider a symmetric top of mass $m$ with a fixed pivot at the origin \point{O}, as shown in \rfig{symtop_fixed}. 
The inertial frame $XYZ$ has a vertically upward $Z$-axis, with gravity $g$ acting downward. 
The center of mass lies at a distance $l$ from \point{O} along the symmetry axis ($z$-axis), which defines the body-fixed frame $xyz$ aligned with the principal axes.
Taking the alignment of $xyz$ with $XYZ$ as the reference orientation, the body-fixed inertia tensor is given by $I=\diag(I_1,I_1,I_3)$.

\begin{figure}[h]
    \centering
    \includegraphics[scale=0.65]{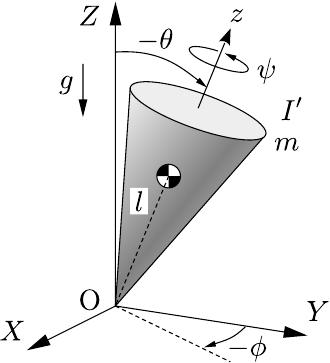}
    \caption{Symmetrical top with one point fixed \cite{Goldstein2011,Landau1976mechanics}.}
    \lfig{symtop_fixed}
\end{figure}

\subsubsection{Kinematics}
\lsec{symtop_fixed:Kinematics}

For this problem, the $ZXZ$ Euler angles $\thh=[\theta^i] =(\phi, \theta, \psi)^T$ are employed
as the orientation of the top, which yields the rotation matrix:
\begin{align}
    R = R(\thh) &:= R_Z(\phi)R_X(\theta)R_Z(\psi)
    \notag\\
    &=
    \begin{bmatrix}
        - \sin\phi  \sin\psi  \cos\theta  + \cos\phi  \cos\psi  & - \sin\phi  \cos\psi  \cos\theta  - \sin\psi  \cos\phi  & \sin\phi  \sin\theta \\
        \sin\phi  \cos\psi  + \sin\psi  \cos\phi  \cos\theta  & - \sin\phi  \sin\psi  + \cos\phi  \cos\psi  \cos\theta  & - \sin\theta  \cos\phi \\
        \sin\psi  \sin\theta  & \sin\theta  \cos\psi  & \cos\theta
    \end{bmatrix}
    ,
    \leqn{koma:R}
\end{align}
where $R_{(\bullet)}$ denotes the rotation matrices about the $(\bullet)$-axes, respectively. 
The angle $\phi$ represents the precession (rotation of the $z$-axis about the $Z$-axis), 
$\theta$ is the nutation (tilt of the $z$-axis), and 
$\psi$ is the spin (rotation of the top about the $z$-axis). 

Substituting the rotation matrix $R$ from \reqn{koma:R} into the formula in \reqn{A}
yields the body-fixed kinematic relation
\begin{equation}
    \omm = A\dthh
    ,\quad 
    A
    = 
    [A^j_i]
    = 
    \left[
      \frac{1}{2}\levi{_a}{^{jc}}{}
      {R^T}^a_b\pd{R^b_c}{\theta^i}
    \right]
    =
    \begin{bmatrix}
        \sin\theta\sin\psi  &  \cos\psi & 0\\
        \sin\theta\cos\psi  & -\sin\psi & 0\\
        \cos\theta          & 0         & 1
    \end{bmatrix}
    ,
    \leqn{koma:A:body}
\end{equation}
where the right pseudoinverse of $A$ (the true inverse in this case), denoted by $B$, yields the inverse kinematic relation
\begin{equation}
    \dthh = B\omm
    ,\quad 
    B = 
    \begin{bmatrix}
        \sin\psi/\sin\theta  & \cos\psi/\sin\theta   & 0 \\
        \cos\psi             & -\sin\psi             & 0 \\
        -\sin\psi/\tan\theta & -\cos\psi/\tan\theta  & 1 \\
    \end{bmatrix}
    .
    \leqn{koma:B:body}
\end{equation}

\subsubsection{Equations of motion derived by the proposed EL equation}

The symmetrical top shown in \rfig{symtop_fixed} is described by the Lagrangian \citep{Goldstein2011}
\begin{equation}
    \clL = \Kine - \Pote 
    = \frac{1}{2}\omm^T I \omm - mgl\cos\theta
    \leqn{fixed:Lag}
    .
\end{equation}
Substituting this Lagrangian into the proposed EL equation \reqn{modELE:vector:body} with $\QQnc=\bm 0$
directly yields the equations of motion
\begin{align}
    & 
    \OD{}{t}\Pbk{\pd{\clL}{\omm}} + \omm\times\pd{\clL}{\omm} - B^T\pd{\clL}{\thh}
    = 
    \bm0
    ,
    \notag\\
    &
    \pd{\clL}{\omm}
    =
    \bmat{I_1\omega^1\\ I_1\omega^2\\ I_3\omega^3}
    ,\quad
    \od{}{t}\Pbk{\pd{\clL}{\omm}} 
    =
    \bmat{I_1\dot\omega^1\\ I_1\dot\omega^2\\ I_3\dot\omega^3}
    ,\quad
    B^T\pd{\clL}{\thh}
    =
    mgl\sin\theta
    \begin{bmatrix}
         \cos\psi \\
        -\sin\psi \\
         0        \\
    \end{bmatrix}
    .
    \leqn{koma:ELEom}
\end{align}
Solving \reqn{koma:ELEom} simultaneously with \reqn{koma:B:body}
yields the top's motion $\XX(t) = (\thh(t)^T,\omm(t)^T)^T$.

\subsubsection{Consistency with the classical Euler equation}

While trivial from the rigorous discussion in \rsec{classical:Euler},
we show that the resulting equations \reqn{koma:ELEom} coincides the classical Euler equation defined by
\begin{equation}
    I \domm + \omm\times(I\omm) = \tauu,    
    \leqn{koma:EulerEOM}
\end{equation}
where $\tauu$ denotes the body-fixed components of the external torque.
In this example, $\tauu$ is exclusively generated by gravity force
\begin{equation}
 \FF_g = -mg\eeZ, \quad \eeZ:=(0, 0, 1)^T   
\end{equation}
which generates the torque in the reference frame $XYZ$:
\begin{equation}
    \tauu' 
    = \Pbk{
        l\cdot R_Z(\phi) R_X(\theta)\eeZ
    }\times
    \FF_g
    =
    mgl\sin\theta
    \bmat{
        \cos\phi\\
        \sin\phi\\
        0
    }
    .    
    \leqn{koma:T}
\end{equation}
Transforming this into the body-fixed components, we have
\begin{equation}
    \tauu = R^T\tauu' 
    =    
    mgl
    \begin{bmatrix}
         \sin\theta\cos\psi \\
        -\sin\theta\sin\psi \\
         0        \\
    \end{bmatrix}
    =    
     mgl\sin\theta
    \begin{bmatrix}
         \cos\psi \\
        -\sin\psi \\
         0        \\
    \end{bmatrix}
    .
    \leqn{koma:T'}
\end{equation}
Substituting \reqn{koma:T'} into the Euler equation \reqn{koma:EulerEOM}
yields a result which agrees with \reqn{koma:ELEom} derived from our proposed EL equation.

\subsection{Symmetrical top suspended by a spring}
\lsec{symtop_spring}

The proposed EL formulation is further verified by demonstrating the coupled translational and rotational motion of a symmetric top suspended by a spring, as shown in \rfig{symtop_spring}.
This top is identical to the one shown in \rfig{symtop_fixed}, except that its pivot $P$ is suspended from the origin $\point{O}$ by a massless spring with spring constant $k$ and zero natural length. Let us denote the spatial displacement of the center of gravity (COG) of the top by $\XX := (X^1,X^2,X^3)^T$, and the spring force acting on the pivot by $\FF_k:=(F^1_k,F^2_k,F^3_k)^T$.
Other specifications and conditions are identical to those for the top in \rsec{symtop_fixed}.

\begin{figure}[h]
    \centering
    \includegraphics[scale=0.65]{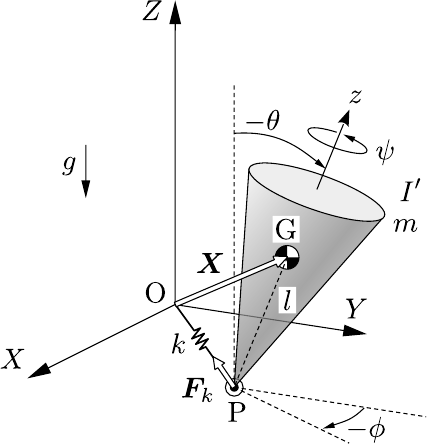}
    \caption{Symmetrical top suspended by a spring.}
    \lfig{symtop_spring}
\end{figure}

\subsubsection{Kinematics}

We employ the same Euler angles $\thh$, rotation matrix $R(\thh)$, and kinematic matrices $A$ and $B$ as those given in \rsec{symtop_fixed:Kinematics}.
The configuration of this system is described by the generalized coordinates
\begin{equation}
    \qq := (\XX^T, \thh^T)^T
    = (X^1,X^2,X^3,\phi,\theta,\psi)^T
    ,
\end{equation}
where $\XX$ and $\thh$ denote the COG displacement and the Euler angles, respectively.

Furthermore, we consider the displacement vector of the pivot $\point{P}$ from the origin $\point{O}$:
\begin{align}
    &
    \PP
    :=
    \varrow{OP}
    = \XX + \varrow{GP}
    ,\quad
    \varrow{GP}
    := -lR(\thh)\eeZ
    ,\quad
    \eeZ := (0,0,1)^T
    .
    \leqn{PP}
\end{align}

\subsubsection{Equations of motion derived by the proposed method}

The kinetic energy $\Kine$ and the potential energy $\Pote$ are given by
\begin{equation}
    \Kine 
    =
    \frac{1}{2}\omm^TI\omm 
    + \frac{m}{2}\dot\XX^T\dot\XX
    ,\quad
    \Pote
    =
    \frac{k}{2}\PP^T\PP + mgX_3
    ,
\end{equation}
where
\begin{align}
    \PP^T\PP
    &
    =(\XX-lR(\thh)\eeZ)^T(\XX-lR(\thh)\eeZ)
    \notag\\&
    =(\XX^T-l\eeZ^TR^T(\thh))(\XX-lR(\thh)\eeZ)
    \notag\\&
    = \XX^2 - 2l\XX^TR(\thh)\eeZ + l^2\eeZ^TR^T(\thh)R(\thh)\eeZ
    = \XX^2 - 2lR(\thh)\eeZ\cdot\XX + l^2
    ,
\end{align}
where ``$\cdot$'' denotes the inner product.  
This yields the Lagrangian of this coupled translational and rotational motion:
\begin{equation}
    \clL = \Kine - \Pote
    =
    \frac{1}{2}\omm^TI\omm + \frac{m}{2}\dot\XX^2
    -\frac{k}{2}(\XX^2 - 2lR(\thh)\eeZ\cdot\XX + l^2) - mgX_3
    .
\end{equation}

Since this problem is a specific case of the proposed general formulation in \reqn{ELE:general},
we apply it by setting $\qq=\XX$ and $\QQnc_{\thh}=\QQnc_{\xx}=\bm{0}$, 
which yields the following set of EL equations
\begin{align}
    &
    \od{}{t}\Pbk{\pd{\clL}{\omm}}
    +
    \omm\times \pd{\clL}{\omm}
    -
    B^T\pd{\clL}{\thh}
    =
    \bm 0
    ,
    \\
    &
    \od{}{t}\Pbk{\pd{\clL}{\dot\XX}}
    -
    \pd{\clL}{\XX}
    =
    \bm 0
    .
\end{align}
These provide the equations of motion for translation and rotation as
\begin{equation}
    \begin{dcases}
        m\ddot\XX 
        = 
        \FF_{\omm} 
        :=
        \pd{\clL}{\XX}
        ,
        &\quad\text{(translation)}         
        \\
        I\domm + \omm\times(I\omm) 
        = 
        \tauu_{\omm}
        :=
        B^T\pd{\clL}{\thh}
        ,
        &\quad\text{(rotation)} 
    \end{dcases}
    \leqn{symtopspring:ELEom}
\end{equation}
where the torque $\tauu_{\omm}$ and force $\FF_{\omm}$ are evaluated as
\begin{align}
    & \FF_{\omm}
    := \pd{\clL}{\XX}
    = - k\XX + kl R(\thh)\eeZ - mg\eeZ
    ,
    \leqn{force:ELEom}
    \\
    & \tauu_{\omm}
    := B^T\pd{\clL}{\thh}
    = kl B^T\pd{(R(\thh)\eeZ\cdot\XX)}{\thh}
    .
    \leqn{torque:ELEom}
\end{align}

\subsubsection{Consistency with the Newton--Euler equations}

The derivation of equations of motion is also demonstrated via the classical Newtonian approach,
directly based on the Newton--Euler equations \cite{Goldstein2011,Williams1996applied},
\begin{equation}
    m\ddot\XX = \FF
    ,\quad 
    I\domm + \omm\times(I\omm) = \tauu
    ,
    \leqn{symtopspring:Newton-Euler}
\end{equation}
where $\XX$ and $\FF$ denote the reference COG displacement vector and external force vector,
$\omm$ and $\tauu$ represent the body-fixed angular velocity vector and external torque vector,
and $m$ and $I$ are the corresponding mass and inertia tensor, respectively.

The force $\FF$ and torque $\tauu$ in \reqn{symtopspring:Newton-Euler} are geometrically derived as follows.
The displacement vector $\PP$ in \reqn{PP} yields the spring force
\begin{equation}
    \FF_k = -k\PP = -k\Pbk{\XX - l R(\thh)\eeZ}
    .  
\end{equation}
Hence, the net force $\FF$ acting on the top is obtained as 
the sum of the spring force $\FF_k$ and the gravitational force $\FF_g := -mg\eeZ$:
\begin{equation}
    \FF = \FF_k + \FF_g = -k\Pbk{\XX - l R(\thh)\eeZ} - mg\eeZ
    ,
    \leqn{force:Euler}
\end{equation}
and the net torque about the COG is given by
\begin{equation}
    \tauu' 
    = \varrow{GP}\times\FF_k
    = -l R(\thh)\eeZ\times\Pbk{-k\Pbk{\XX - l R(\thh)\eeZ}}
    = kl(R(\thh)\eeZ)\times\XX
    ,
\end{equation} 
whose body-fixed representation is expressed as
\begin{equation}
    \tauu = R(\thh)^T\tauu' 
    = kl\eeZ\times(R(\thh)^T\XX)
    .
    \leqn{torque:Euler}
\end{equation}

We can now demonstrate that the equations of motion \reqn{symtopspring:ELEom} obtained via our proposed EL framework 
agree with the Newton--Euler equations \reqn{symtopspring:Newton-Euler}, 
provided that $\FF_{\omm} = \FF$ and $\tauu_{\omm} = \tauu$ hold in \reqn{force:ELEom}, \reqn{torque:ELEom}, \reqn{force:Euler}, and \reqn{torque:Euler}, respectively.

Regarding the forces, the relation $\FF_{\omm} = \FF$ clearly holds upon comparison of \reqn{force:ELEom} and \reqn{force:Euler}. 
As for the torques, $\tauu_{\omm} = \tauu$ is verified by comparing their respective components. Both torques are linear with respect to $\XX$, and their matrix representations coincide as follows:
\begin{align}
    &
    \tauu_{\omm} = kl H\XX = \tauu
    ,
    \notag\\&
    H :=
    \begin{bmatrix}
        \sin{\phi} \cos{\theta} \cos{\psi} + \sin{\psi} \cos{\phi} 
        & \sin{\phi} \sin{\psi} - \cos{\phi} \cos{\theta} \cos{\psi} 
        & - \sin{\theta} \cos{\psi}\\
        - \sin{\phi} \cos{\theta} \sin{\psi} + \cos{\phi} \cos{\psi} 
        & \sin{\phi} \cos{\psi} + \cos{\phi} \cos{\theta} \sin{\psi}  
        & \sin{\psi} \sin{\theta}\\
        0 & 0 & 0
    \end{bmatrix}
    .
\end{align}

Accordingly, our proposed EL formulation is shown to be consistent with the classical Newton--Euler equations 
for the coupled translational and rotational motion of the top.
This result demonstrates that the advantages of our formulation asserted in \rsec{classical:Euler} persist even for the coupled translational and rotational problem considered in this section. 
Specifically, the explicit force--torque analysis 
required to determine $\FF$ in \reqn{force:Euler} and $\tauu$ in \reqn{torque:Euler} is bypassed; 
as with the standard EL framework, 
simply defining the potential energy automatically handles the necessary coordinate transformations.

\newpage
\section{Conclusion}

In this study, we have proposed a novel variant of the Euler--Lagrange (EL) equation explicitly formulated 
in terms of the body-fixed angular velocity vector $\omm$ and orientation parameters $\thh$. 
By analyzing the stationarity condition of the action integral via the calculus of variations and tensor analysis, 
we have established a modified EL framework that seamlessly addresses spatial rigid body dynamics. 
The key theoretical findings and modeling advantages of the proposed method are summarized as follows:

\begin{enumerate}
  \item \textbf{Consistency with classical rotational dynamics:} For rigid body rotational motion subject to potential forces, the proposed EL equation directly yields equations of motion that are fully consistent with the classical Euler equations.
  \item \textbf{Simplification of mechanical modeling:} Unlike the classical Euler approach, which relies on explicit geometry-based force--torque analysis, the proposed formulation inherently incorporates potential gradients and coordinate transformations within the unified Lagrangian framework.
  \item \textbf{Seamless integration of coupled dynamics:} As demonstrated by the case study of a symmetrical top suspended by a spring, combining the proposed rotational EL equation with the conventional EL equation for translation provides a straightforward formulation for coupled translational and rotational motion, exhibiting exact agreement with the Newton--Euler equations.
  \item \textbf{Identification of structural discrepancy in the standard EL approach:} We have demonstrated that applying direct velocity coordinate transformations to the standard EL equation introduces an inherent discrepancy tensor, $\Delta_{\mathrm{EL}} := (\dot{A}B)^T - [\omm\times]$, causing the resulting equations to deviate systematically from both the classical Euler equation and our proposed formulation.
\end{enumerate}

The proposed framework retains the systematic convenience of Lagrangian mechanics while recovering the physical clarity of the classical Euler equations. Consequently, it offers a versatile and intuitive modeling tool for spatial rigid body dynamics across physics and engineering applications.

As a natural continuation of this work, future research will investigate the physical origin and geometric interpretation of the discrepancy $\Delta_{\mathrm{EL}}$ inherent in the standard EL approach, particularly through comparison with the Hamiltonian formulation.

\section*{Acknowledgment}

The author expresses his deep gratitude to Associate Professor Takahiro Yajima of Utsunomiya University for his invaluable guidance on tensor analysis. 
The author is also grateful to Naoto Tokuyama, a graduate student at Utsunomiya University, for pointing out typographical errors in the manuscript.


\end{document}